\documentclass[sigconf]{acmart}

\usepackage{multirow}
\usepackage{mathtools}
\usepackage{soul}

\acmConference[SPLC’22]{26th ACM International Systems and Software Product Lines Conference}{12-16 September, 2022}{Graz, Austria}

\title{Adaptive Behavioral Model Learning for Software Product Lines}

\author{Shaghayegh Tavassoli}
\email{sh.tavassoli@ut.ac.ir}
\orcid{0000-0002-2474-7390}
\affiliation{%
  \institution{University of Tehran}
  \city{Tehran}
  \country{IR}
}

\author{Carlos Diego N. Damasceno}
\email{d.damasceno@cs.ru.nl}
\orcid{0000-0001-8492-7484}
\affiliation{%
  \institution{Radboud University Nijmegen}
    \city{Nijmegen}
  \country{NL}
}

\author{Ramtin Khosravi}
\orcid{0000-0001-6393-0959}
\email{r.khosravi@ut.ac.ir}
\affiliation{%
  \institution{University of Tehran}
  \city{Tehran}
  \country{IR}
}

\author{Mohammad Reza Mousavi}
\orcid{0000-0002-4869-6794}
\email{mohammad.mousavi@kcl.ac.uk}
\affiliation{%
  \institution{King's College London}
  \city{London}
  \country{UK}
  \postcode{WC2R 2LS}
}

\copyrightyear{2022}
\acmYear{2022}
\setcopyright{acmcopyright}\acmConference[SPLC '22]{26th ACM International Systems and Software Product Line Conference - Volume A}{September 12--16, 2022}{Graz, Austria}
\acmBooktitle{26th ACM International Systems and Software Product Line Conference - Volume A (SPLC '22), September 12--16, 2022, Graz, Austria}
\acmPrice{15.00}
\acmDOI{10.1145/3546932.3546991}
\acmISBN{978-1-4503-9443-7/22/09}

\begin{document}

\begin{abstract}
Behavioral models enable the analysis of the functionality of software product lines (SPL), e.g., model checking and model-based testing. Model learning aims to construct behavioral models. Due to the commonalities among the products of an SPL, it is possible to reuse the previously-learned models during the model learning process. In this paper, an adaptive approach, called $\text{PL}^*$, for learning the product models of an SPL is presented based on the well-known $L^*$ algorithm. In this method, after learning each product, the sequences in the final observation table are stored in a repository which is used to initialize the observation table of the remaining products. 
The proposed algorithm is evaluated on two open-source SPLs and the learning cost is measured in terms of the number of rounds,  resets, and  input symbols. The results show that for complex SPLs, the total learning cost of $\text{PL}^*$ is significantly lower than that of the non-adaptive  method in terms of all three metrics. Furthermore, it is observed that the order of learning  products affects the efficiency of  $\text{PL}^*$. We introduce a heuristic to determine an ordering which reduces the total cost of adaptive learning.
\end{abstract}



\maketitle

\section{Introduction}
Models are the foundations of many rigorous analysis techniques in engineering in general and software engineering in particular. Behavioral models specify how a system behaves as a result of interacting with its user and environment. Examples of behavioral models include variants of state machines and sequence diagrams. Behavioral models are often non-existent or outdated and one needs to reconstruct them from implementations in order to enable further analysis \cite{DBLP:journals/ese/DamascenoMS21}. Model learning is a mechanized approach that comes to rescue in such situations \cite{DBLP:journals/cacm/Vaandrager17}.

In software product lines, model learning is challenged by variability \cite{DBLP:journals/tse/GalsterWTMA14,DBLP:conf/wicsa/GurpBS01}: one needs to learn behavioral models over the variability space and if performed crudely, this can be practically impossible. The key to overcome this challenge is to reuse the learned models and their underlying data structures while moving across the variability space \cite{DBLP:books/daglib/0018329}. Adaptive model learning \cite{groce_2002_amc} is fit for this purpose, because its algorithms are precisely designed to reuse the results of the past queries, as well as the structure of the behavioral models in the subsequent learning process \cite{DBLP:conf/ifm/DamascenoMS19}.

In this paper, we design an adaptive learning algorithm  for software product lines, called $PL^*$, and evaluate its efficiency against its non-adaptive counterparts. Two important components of a model learning algorithm are the membership queries (checking for the output to a given sequence of inputs) and the equivalence queries (verifying the model learned hitherto). The main factors in evaluating the efficiency of a model learning algorithm are the number of learning rounds (and equivalence queries), the number of resets, and the total number of input symbols used in learning  \cite{DBLP:journals/iandc/Angluin87,DBLP:journals/cacm/Vaandrager17,DBLP:conf/ifm/DamascenoMS19}.
We evaluate the efficiency of our algorithm on two subject systems. We statistically evaluate our results (with 3-wise sampling, based on earlier experiments \cite{DBLP:journals/ese/DamascenoMS21}, and with different product orderings) and observe, with high statistical confidence, that $\text{PL}^*$ is more efficient than the non-adaptive approach. Hence, we affirmatively answer the following research questions:

\begin{enumerate}
\itemsep0em
         \item[RQ1] Does adaptive learning lead to fewer learning rounds and equivalence queries?
         \item[RQ2] Does adaptive learning lead to fewer resets?
         \item[RQ3] Does adaptive learning lead to fewer total number of input symbols?
\end{enumerate}
       
In standard (non-adaptive) model learning, the order of learning the products is immaterial, since no information is brought forward to learning the next products. In our adaptive learning method, we observe a stark difference in terms of efficiency among different orders. We define and formalize a heuristic to provide an efficient product ordering in the learning process and statistically establish a correlation between our proposed ordering and the efficiency of the learning algorithm (in terms of the total number of resets and the total number of input symbols) as the answer to our last research question: 

\begin{enumerate}
\itemsep0em
         \item[RQ4] How does the choice of product ordering influence the efficiency of the learning process?
\end{enumerate}  

To our knowledge, this is the first application of adaptive model learning to software product lines (we refer to Section 
\ref{sec:related} for an analysis of the related work). It is a first step in this direction, which will pave the way for a line of research extending various model learning techniques to a family-based approach. This natural extension would require parameterizing the data structure we use in our approach (called observation tables \cite{DBLP:journals/iandc/Angluin87}) with feature expressions.

The rest of this paper is organized as follows. In Section \ref{sec:related}, we review the related work and position our research within the broader fields of model learning and software product lines. In Section \ref{sec:back}, we recall some basic concepts and definitions from the aforementioned fields. In Section \ref{sec:adaptive}, we present our adaptive model learning algorithm. In Section \ref{sec:empirical}, we outline our empirical evaluation methodology and describe the design of our experiments. In Section \ref{sec:results}, we discuss the results of our experiments and reflect on the threat to the validity of our results. In Section \ref{sec:conc}, we conclude the paper and present the directions of our future research. A package containing the source code, models, and test scripts is available at \url{https://github.com/sh-t-20/artifacts}

\section{Related Work}\label{sec:related}
In this section, we review the related work in three broad areas:  
adaptive model learning \cite{groce_2002_amc,Chaki2008,Windmuller2013_ACQ,huistra2018_adaptlearning,DBLP:conf/ifm/DamascenoMS19,yang2019_activepassivelearning},
machine learning in SPLs \cite{PEREIRA2021111044,DBLP:conf/splc/DamascenoMS19,DBLP:journals/ese/DamascenoMS21,lesoil2021_deepswvary}, and
feature model mining \cite{Haslinger2011_refm,Ryssel2011_ExtractionFM,almsiedeen2014_refm_fca}.

\subsubsection*{Adaptive Model Learning}
Adaptive model learning \cite{groce_2002_amc} is an extension of traditional model learning by reusing pre-existing models.
Groce, Peled  \& Yannakakis \cite{groce_2002_amc} are among the first to reuse inaccurate models for adaptive model learning and model checking. 
The authors' results suggest that adaptive learning is especially useful when model updates are led by small changes with limited impact \cite{groce_2002_amc}. 
Our results corroborate their observation in the setting of software product lines; 
namely, we show that adaptive learning is more  efficient when the order of products comprises fewer new non-mandatory features added in each step. 
Windm\"uller et al. \cite{Windmuller2013_ACQ} show that adaptive learning can be used to periodically build models from evolving complex applications.
Also, they show that reusing separating sequences derived from models of previous versions can steer the learning process to find maintained states \cite{Windmuller2013_ACQ}.
Huistra, Meijer,  \& van de Pol \cite{huistra2018_adaptlearning} report that the performance of adaptive learning is influenced by the SUL's complexity, the size of its update, and the  quality of suffixes.
Additionally, the authors report evidence that, if a set of reused separating sequences has low state distinguishing capacity, then irrelevant queries should be expected \cite{huistra2018_adaptlearning}. 
Chaki, Clarke, Sharygina  \& Sinha \cite{Chaki2008} presented an approach for efficiently model checking software upgrades by revalidating sequences from reused observation tables  \cite{Chaki2008}. 
More recently, Damasceno et al. \cite{DBLP:conf/ifm/DamascenoMS19} showed that existing adaptive learning techniques are prone to performance issues when there are large differences between the reused and updated model.
To address this issue, they introduced a novel adaptive learning technique that gradually revalidates and reuses sequences and outperforms state-of-the-art adaptive learning techniques \cite{DBLP:conf/ifm/DamascenoMS19}.
Our results crucially build upon these earlier results and bring them to a new domain: 
software product lines provide a specific paradigm for adaptive learning, where the choice of adaptations can be controlled by the product sampling order. 
Our work differs from these by focusing on the reuse of multiple observation tables in an observation table repository and is to our knowledge the first attempt to lift adaptive model learning to the scope of software families.
Yang et al. \cite{yang2019_activepassivelearning} present a way to combine the results of passive- and active model learning \cite{yang2019_activepassivelearning}. Our work differs from this piece of work in that we consider active model learning; the combination of adaptive active and passive learning for product families is a promising area for future work.

\subsubsection*{Machine and Model Learning in SPLs} 
Several studies applied machine learning \cite{PEREIRA2021111044,DBLP:journals/ese/TemplePABJR21} and model learning \cite{DBLP:conf/splc/DamascenoMS19,DBLP:journals/ese/DamascenoMS21} in software product lines. 
For an extensive literature review on machine learning applied to SPLs, we refer the interested reader to Pereira et al. \cite{PEREIRA2021111044}.
Lesoil \cite{lesoil2021_deepswvary} have recently showed that \textit{variability} can be present at multiple layers of a system (e.g., at the hardware, software, and input data levels) and raise concerns about challenges in the adoption of machine learning principles in variability analysis.
Family-based modeling approaches have been developed to enable efficient model-based testing of SPLs without exhaustively going through each and every product.
Nevertheless, the creation and maintenance of family models are still difficult and time-consuming \cite{oster2012_feat}. 
To  mitigate this, Damasceno et al. \cite{DBLP:conf/splc/DamascenoMS19,DBLP:journals/ese/DamascenoMS21} introduced family model learning as a means for building behavioural variability models for SPLs. 
Using a benchmark set of 105 product models, the authors showed that succinct family models can be learned by matching and merging state machine models \cite{DBLP:journals/ese/DamascenoMS21}, particularly when there is a high degree of reuse among the SULs \cite{DBLP:conf/splc/DamascenoMS19}. 
Additionally, they show that feature coverage criteria (particular, up to 3-wise) can alleviate the costs of learning family model by sampling product sets that cover the behavior of product families.
These results have sparked the interest of the SPL community in pursuing further investigations at the intersection of model learning and variability analysis \cite{fortz2021_lifts}.
Our work advances this research line by reusing observation tables across multiple products. Particularly, using adaptive learning techniques is a novelty of our approach. 

\subsubsection*{Feature Model Mining}
Feature models are a key asset in variability management and analysis \cite{benavides2010_afm}.
Using SAT- \cite{leberre2012_sat4j} or SMT \cite{sprey_smt-based_2020} solvers, feature models are amenable to automated reasoning.
However, as an SPL may also be built using extractive and reactive approaches \cite{apel2013_fosd_ch2}, SPL projects may initially lack feature models  \cite{Haslinger2011_refm}. To address this issue, reverse engineering concepts have been used to (semi-)automate the construction of feature models from sets of product configurations \cite{Haslinger2011_refm,Ryssel2011_ExtractionFM,almsiedeen2014_refm_fca}. Our work complements the role of such structural variability model learning techniques (aka feature model mining) by providing an efficient means to extract behavioral variability models. The integration of these two sets of techniques is a promising line of future research. 

\section{Background}\label{sec:back}

In this section, some of the terms used in this paper are described. The software product lines are briefly explained. Some notations for modeling an SPL and its products are defined. The non-adaptive model learning process is explained. Some of the metrics used for evaluating the efficiency of model learning are defined. Also in this section, the product sampling concept is briefly described.

\subsection{Software Product Lines}

\begin{figure}[!ht]
    \centering
    \includegraphics[trim=0 0.25cm 0 0,scale=0.7]{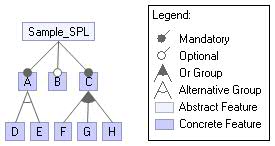}
    \caption{The feature model of a sample SPL}
    \label{fig:feature_model}
\end{figure}

A software product line (SPL) is a set of software products that have a common set of features and are designed for a specific requirement \cite{DBLP:books/daglib/0019719}. An SPL is defined by a set of features $F$ and a feature model \cite{DBLP:journals/ese/DamascenoMS21,DBLP:conf/splc/DamascenoMS19}. A feature-model \cite{kang1990feature,DBLP:conf/re/SchobbensHT06} is a structural variability model representing the hierarchical structure of the SPL features. Each product $p$ consists of a subset of the SPL features. The set of valid product configurations is specified by the feature model. In this representation, features are classified into mandatory and optional types. Mandatory features are present in all valid configurations by default. From a group of \textit{alternative} features, only one of them can be present in each product. When a set of features are defined using \textit{or}, each product may contain one or more of them \cite{DBLP:journals/ese/DamascenoMS21,kang1990feature}. Figure \ref{fig:feature_model} shows the feature model of a sample SPL. In this figure, \textit{A} and \textit{C} are mandatory features and \textit{B} is an optional feature. The features \textit{D} and \textit{E} are alternatives. \textit{F}, \textit{G} and \textit{H} form an `or' group of features. This SPL contains 28 valid product configurations. 

\subsection{Finite State Machines}

A finite state machine (FSM) \cite{gill1962introduction} is a widely used behavioral model which is defined as the tuple $M=\langle S,s_{0},I,O,\delta,\lambda\rangle$. In this definition, $S$ is the set of states and $s_{0}$ is the initial state ($s_{0}\in S)$. The set of input alphabet is represented by $I$ and the set of outputs is denoted by $O$. The transition function $\delta$ determines the next state, $s_{2}\in S$, assuming that the FSM is in state $s_{1}\in S$ and the input $a\in I$ is presented ($\delta (s_{1},a)=s_{2}$). The output function is represented by $\lambda$ which is a mapping from a pair of a state and an input to an output. The state machines learned by the existing model learning methods are deterministic FSMs. In a deterministic FSM, for each state and input alphabet, there is at most one transition and one output \cite{DBLP:conf/ifm/DamascenoMS19,DBLP:journals/cacm/Vaandrager17}.

\subsection{Model Learning}

Model learning \cite{DBLP:journals/cacm/Vaandrager17} is a method used to construct the behavioral model of a software system in the form of a state machine. Model learning is classified into two types: passive  and active. In passive learning, different runs of software (e.g., log files) are used for learning a behavioral model of the software. In active learning, however, a model is learned by interacting with the system under learning (SUL) through various types of queries and observing the resulting outputs. The $L^{*}$ algorithm \cite{DBLP:journals/iandc/Angluin87} proposed by Dana Angluin, is a seminal example of active model learning, where two types of queries are used: a membership query (MQ) is used to determine the output sequence for a given input sequence. An equivalence query (EQ) is used to ask if the constructed hypothesis $H$ is language-equivalent to the SUL. The query results are stored in an observation table (described below), and the learning is performed in rounds, in each of which a hypothesis is constructed \cite{DBLP:journals/iandc/Angluin87,DBLP:conf/ifm/DamascenoMS19,DBLP:journals/cacm/Vaandrager17}.  

\subsubsection{Observation Table}

An observation table is defined as a triple $\mathit{OT}= (S,E,T)$, where $S\subseteq I^{*}$ is a finite prefix-closed set of prefixes (transfer sequences); $E\subseteq I^{+}$ is a finite set of suffixes (separating sequences) and $T:  I^{+} \times I^{+} \rightarrow I^+$ is a function such that for each $s \in S \cup S.I$ and $e \in E$, $T(s,e)$ is the SUL's output suffix of size $|e|$ for the  $s.e$ input sequence. An observation table can be represented as a 2-dimensional array, where each row is a  subset of $S.I$ with a representative $s \in S.I$,  and each column is a sequence $e \in E$. An observation table is \textit{closed} if for all $s_{1}\in S.I$, there exists a prefix $s_{2}\in S$ such that $\mathit{row}(s_{1})$ equals $\mathit{row}(s_{2})$. An observation table is \textit{consistent} if for all $s_{1},s_{2}\in S$ such that $\mathit{row}(s_{1})=\mathit{row}(s_2)$, $\mathit{row}(s_{1}.v)$ equals $\mathit{row}(s_{2}.v)$ for all $v\in I$ \cite{DBLP:journals/iandc/Angluin87,DBLP:conf/ifm/DamascenoMS19,DBLP:journals/cacm/Vaandrager17}.

\subsubsection{The $L^{*}$ Algorithm}

In this section, Angluin's $L^{*}$ algorithm \cite{DBLP:journals/iandc/Angluin87} is described. At the beginning of this algorithm, sets $S$ and $E$ are initialized to $\{\epsilon\}$ and the initial $T$ values are obtained by posing MQs. The following steps are repeated until the observation table is consistent and closed:
\begin{itemize}
    \item If the observation table is not consistent, the algorithm finds $s_{1},s_{2}\in S$, $v\in I$ and $e_{1}\in E$ such that $\mathit{row}(s_{1})=\mathit{row}(s_{2})$ and $T(s_{1}.v,e_{1})\neq T(s_{2}.v,e_{1})$. Then, $v.e_{1}$ is added to $E$ and the new values of $T$ are calculated by posing MQs.
    \item If the observation table is not closed, the algorithm finds $s_{1}\in S$ and $v\in I$ such that $\mathit{row}(s_{1}.v)\neq \mathit{row}(s)$ for all $s\in S$. Then, $s_{1}.v$ is added to $S$ and the new values of $T$ are obtained using MQs.
\end{itemize}
When the observation table is closed and consistent, the algorithm constructs a hypothesis $H$ and poses an EQ to verify it. If $H$ is correct, the learning algorithm terminates. If the hypothesis $H$ is not correct, a counterexample is provided. A counterexample is an input sequence in which the result of $H$ is different from the result of the SUL \cite{DBLP:journals/iandc/Angluin87,DBLP:conf/ifm/DamascenoMS19,DBLP:journals/cacm/Vaandrager17}.
Then, the counterexample is used to update the observation table by adding prefixes or suffixes (for which several heuristics have been proposed) \cite{DBLP:journals/ac/IrfanOG13}.

The $L_{M}^{*}$ \cite{DBLP:conf/fm/ShahbazG09} is an active model learning algorithm for learning mealy machines using the settings of $L^{*}$. In $L_{M}^*$, the observation table is defined as $\mathit{OT}=\{S_{M},E_{M},T_{M}\}$ and is initialized using $S_{M}=\{\epsilon\}$ and $E_{M}=I$ \cite{DBLP:conf/fm/ShahbazG09}. In this paper, the $L_{M}^{*}$ algorithm is used to perform the experiments.

\subsection{Product Sampling}

Product-based behavioral analysis of an SPL, can be costly due to the exponential number of valid configurations. Using sample-based approaches may result in increasing the efficiency of the SPL analysis. In these approaches, a subset of valid products is used to cover the behavior of an SPL. Products whose behavior has already been covered by other products are not included in the sample \cite{DBLP:journals/csur/ThumAKSS14}. 
The T-wise \cite{DBLP:conf/models/JohansenHF11} method is one of the sampling techniques applicable in the SPL context. In this method, valid combinations of T-features are used to cover the T-wise interactions of features in the SPL \cite{DBLP:conf/models/JohansenHF11,DBLP:conf/icst/PerrouinSKBT10,DBLP:journals/ese/DamascenoMS21}.

\section{The PL* algorithm}\label{sec:adaptive}

In adaptive model learning, the transfer sequences and the separating sequences in the observation tables of the existing models are reused to initialize the observation table of the new model \cite{DBLP:conf/ifm/DamascenoMS19}. In our approach, we build upon an adaptive learning algorithm \cite{DBLP:conf/ifm/DamascenoMS19}, and apply to learn the FSM models of a set of products sampled from an SPL. Assume $\mathit{Sample}= (p_{1},p_{2},\dots ,p_{n})$ is a sequence of $n$ products sampled from an SPL.

In this algorithm, $\mathit{otRepository}$ is defined as a set of observation tables learned from earlier products. To refer to these observation tables, we use the notation $\mathit{OT}_{i}=(S_{i},E_{i},T_{i})$ for the observation table of product $i$. At the beginning of the learning process, the $\mathit{otRepository}$ is empty. The $\text{PL}^*$ algorithm consists of the following steps:

\begin{enumerate}
\item First, the FSM of $p_{1}$ is learned using a non-adaptive learning method (e.g., using the $L^{*}_M$ algorithm). We only deviate from $L^*$ by initialising the initial set of suffixes (i.e., $E_{1}$) with the alphabet of the product $p_1$. (We do the same for all other subsequent products, as well, i.e., we add their alphabet to their initial set of suffixes.) 

\item Once the $L^*_M$ algorithm successfully terminates, the resulting observation table $\mathit{OT}_{1}$ is added to the $\mathit{otRepository}$. Therefore, $\mathit{otRepository}$ equals $\{\mathit{OT}_{1}\}$. \\

For each $i \in \{2, \ldots, n\}$, the following steps are iteratively repeated: \\

\item The model of product $p_{i}$ is learned using adaptive $L^*_M$. In this step, the observation table of $p_{i}$ is initialized using $\mathit{otRepository}=(\mathit{OT}_{1}, \ldots, \mathit{OT}_{i-1})$.
A sequence is ``defined in the alphabet of $p_{i}$'' if it solely comprises input symbols in the alphabet of $p_{i}$.
To initialize the $\mathit{OT}_{i}$, the set of sequences in $\bigcup_{j\in \{1, \ldots, i-1\}}$ $S_{j}$ which are defined in the alphabet of $p_{i}$, is considered as the initial value of $S_{i}$.
The set of sequences in $\bigcup_{j\in \{1, \ldots, i-1\}}$ $E_{j}$ which are defined in the alphabet of $p_{i}$, are added to $E_{i}$ (note that the alphabet of $p_i$ is initially added to $E_{i}$ by default). 

\item Once adaptive $L^*_M$ terminates, $\mathit{OT}_{i}$ is added to the $\mathit{otRepository}$ and at the end of this step,  $\mathit{otRepository}$ equals $(\mathit{OT}_{1}, \ldots, \mathit{OT}_{i})$.
\end{enumerate}

\begin{figure}[!ht]
    \centering
    \includegraphics[width=0.95\linewidth]{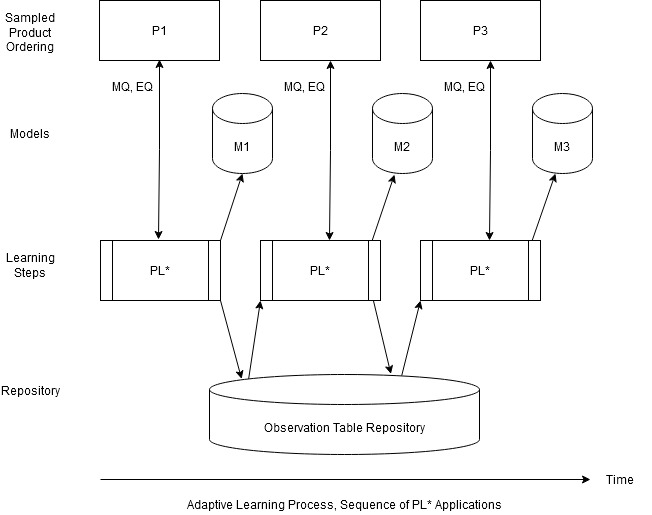}
    \caption{A schematic representation of the proposed adaptive learning method}
    \label{fig:figure_schematic}
\end{figure}

A schematic representation of the proposed adaptive learning method is shown in Figure \ref{fig:figure_schematic}. In this figure, $M_{i}$ is the model learned for the $i$-th product ($p_{i}$). In this figure, arrows from $\text{PL}^*$ processes to the observation table repository show that the observation table of the recently learned product is stored in the repository. The arrows starting from the repository show the sets of sequences in the repository which are used to initialize the observation table of the new products. 
After learning the model of each product, the learned model is incorporated into a family model (a feature-annotated behavioral model of the software product line \cite{DBLP:conf/facs2/FragalSM16}). The two processes of learning the product models and updating the family model can be performed concurrently.

Based on the results of the experiment described in Section \ref{e1}, we observed that the product learning order can affect the efficiency of the $\text{PL}^*$ method.
The product learning order can be random or it can be determined using heuristic methods.
To find a good learning order, it is necessary to determine which characteristics of a learning order result in increasing the efficiency. Based on the results of the experiments (Section \ref{e1}) and observing learning orders with high, medium, and low efficiency, a heuristic is presented to determine the desired learning orders. In the experiments, we observed that when the number of new non-mandatory features that are added by each product is small, the efficiency of the $\text{PL}^*$ method increases. Using this observation, we present a heuristic to provide an ordering which decreases the total cost of learning. 

Suppose the number of non-mandatory features of an SPL is $F$ and a sample of size $n$ from this SPL is available for learning. The product learning order $O=\langle p_{1},p_{2},...,p_{n}\rangle$ is a sequence of products in this sample. If $i$ is smaller than $j$, the FSM of $p_{i}$ must be learned earlier than the FSM of $p_{j}$. The parameter $D$ is defined as follows.
\begin{equation}
\label{eq:1}
    D = \begin{cases*}
    0, & if $F_{i}=0$\\
    \sum_{i=1}^{n} \frac{1}{F_{i}}, & if $F_{i}\ne 0$
    \end{cases*}
\end{equation}
In Equation \ref{eq:1}, $F_{i}$ is the number of new non-mandatory features added by $p_{i}$, i.e., the number of non-mandatory features in $p_i$ not present in any product $p_j$, where $1\leq j < i$.
The reason for using $\frac{1}{F_{i}}$ in this formula is that the added cost of learning decreases as the number of new non-mandatory features increases (i.e., the difference between 1 and $\frac{1}{2}$ is larger than the difference between $\frac{1}{4}$ and $\frac{1}{5}$).

\section{Empirical Evaluation Methodology}\label{sec:empirical}

To evaluate the efficiency of the proposed adaptive learning method, a set of experiments is performed. These experiments are designed to answer the following questions by comparing the quantitative metrics between the $\text{PL}^*$ method and the non-adaptive learning method: 

       \begin{enumerate}
         \item[RQ1] Does adaptive learning lead to fewer total number of rounds and equivalence queries?
         \item[RQ2] Does adaptive learning lead to fewer resets?
         \item[RQ3] Does adaptive learning lead to fewer total number of input symbols?
       \end{enumerate}
These quantitative metrics arise from the way model learning algorithms operate: MQs are the simple and basic building blocks to build a hypothesis about the system under learning and hence, their total number is indicative of how long it takes before the hypotheses are constructed. EQs are much heavier than MQs and their total number heavily influences the performance. In order to put these two types of queries together, one needs to factor in the substantial difference in the size of these two types of queries; this is best achieved by counting the total number of input symbols, which gives us a very natural indicator of the overall performance of the algorithm \cite{DBLP:journals/cacm/Vaandrager17} (RQ3).

While performing these queries, sometimes a reset operation is needed to bring the FSM to a known state and pose further queries. This operation is known to be very costly and is often avoided as much as possible in learning algorithms.
To reset a SUL, it may be necessary to completely restart the system and re-initialize many of its software components. Therefore, performing a reset may take a long time \cite{DBLP:conf/pts/GrozSPO15,DBLP:conf/dagstuhl/GrozSPO16}. 
Hence, we use the total number of resets as another efficiency metric for our comparison (RQ2).

\subsection{Subject Systems}

To evaluate the proposed adaptive learning method, we need access to SPLs with well-defined behavioral (e.g., FSM or labelled transition system) and structural models (e.g., Feature Models and the alphabet of each feature). It must also be possible to obtain the FSM of any valid configuration from these SPLs. The tested SPLs must be complex enough to involve a number of rounds and have a reasonably large number of queries in order to allow for a meaningful comparison. According to the mentioned characteristics, two open-source SPLs are used in the experiments:

\subsubsection{The Minepump SPL}

\begin{figure}[!ht]
    \centering\includegraphics[width=0.9\linewidth,trim=0 0.5cm 0 0]{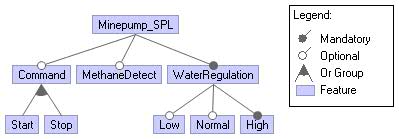}
    \caption{The feature model of the Minepump SPL \cite{classen2010}}
    \label{fig:figure_1}
\end{figure}

The Minepump SPL is presented in \cite{classen2010,classen2011modelling} and is a simple mine-pump controller that includes 9 features (6 non-mandatory features). The feature model of this SPL is shown in Figure \ref{fig:figure_1} \cite{classen2010}. The featured transition system of this SPL is provided in \cite{DBLP:journals/ese/DamascenoMS21}; Using this featured behavioral model, it is possible to obtain the FSM of any valid configuration from this SPL.
Sampling is performed using the 3-wise method. The sample created from this SPL contains 15 products.
The FSMs of the products in this sample have a minimum of 9 states and a maximum of 21 states, and their average number of states is $13.86$.

\subsubsection{The BCS SPL}

The Body Comfort System (BCS) SPL \cite{lity2013delta} is an automotive software system of a Volkswagen Golf model, whose original feature model has 27 features. Each component in this SPL represents a feature and provides a specific functionality. The I/O transition system of each component is provided at \cite{lity2013delta}.

\begin{figure*}
    \centering
    \includegraphics[trim=0 0.35cm 0 0,scale=0.5]{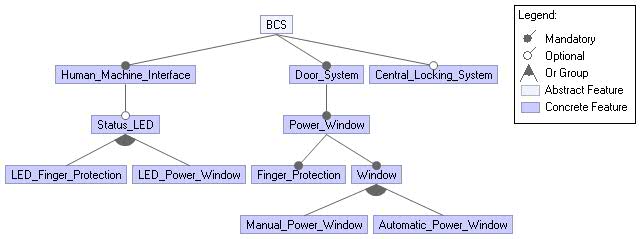}
    \caption{The feature model of a simplified version of the BCS SPL (inspired by \cite{lity2013delta})}
    \label{fig:figure_2}
\end{figure*}

In this paper, a simplified version of the BCS SPL is used. The feature model of the simplified version of the BCS SPL is shown in Figure \ref{fig:figure_2} (taken from \cite{lity2013delta} with minor modifications). The simplified version contains 12 features (6 non-mandatory features).
The product FSMs are constructed using the following steps:
\begin{enumerate}
\item The I/O transition system of each component is converted to a finite state machine (FSM).
\item The FSMs of the components corresponding to the features of each product are merged to construct the FSM of that product.
\end{enumerate}
The sample created from this SPL using the 3-wise method contains 16 products. The FSMs of the products in this sample have a minimum of 14 states and a maximum of 864 states, and their average number of states is $117.25$.

\subsection{Experiment Design}

To perform the experiments, a subset of the valid configurations of each subject SPL is used. The samples are produced using the T-wise product sampling method \cite{DBLP:conf/models/JohansenHF11} and the Chvatal algorithm \cite{DBLP:journals/mor/Chvatal79}.
For sampling using T-wise method, the value of $T$ is set to 3. In \cite{DBLP:journals/ese/DamascenoMS21}, it is shown that in T-wise sampling method, the use of $\text{T}=3$ results in a more precise family model than in cases where $\text{T}=1$ or $\text{T}=2$. In this sampling method, using $T$ greater then 3 is not cost-effective \cite{DBLP:journals/ese/DamascenoMS21}.
Sampling is performed using the FeatureIDE \cite{DBLP:journals/scp/ThumKBMSL14} library. The FSMs of all products in each sample are learned using the $\text{PL}^*$ method and the non-adaptive learning method. The total learning cost is calculated for each learning method.

In these experiments, model learning is performed using the ExtensibleLStarMealyBuilder class of the LearnLib \cite{DBLP:conf/fase/RaffeltS06} library version 0.16.0. In the non-adaptive learning method, before applying  model learning to each product, the observation table is initialized using $S_{M}=\epsilon$ and $E_{M}=I$, where $I$ is the input alphabet of the product \cite{DBLP:conf/fm/ShahbazG09}. In the adaptive learning method, the observation tables are initialized using the method explained in Section 3 (the $\text{PL}^*$ algorithm). Model learning is performed using the following parameters:

\begin{itemize}
    \item The equivalence oracle type is WP, which is an established and structured method for detecting faults (i.e., incorrectly learned states and transitions).
    \item The observation table closing strategy is CloseFirst.
    \item Caching is not used.
\end{itemize}

 To evaluate the experiment results, statistical tests are performed using the SciPy \cite{SciPy} library of Python. The Matplotlib \cite{Hunter:2007} library is used to visualize the results. The experiments show that in the $\text{PL}^*$ method, the order of learning the products affects the total cost of learning. To more accurately evaluate the results, the following experiments are performed:

 \subsubsection{Comparing the Learning Methods}
To compare the efficiency of the $\text{PL}^*$ method with the non-adaptive learning method, a sample of 200 different random learning orders is produced for each subject SPL. Each learning order is considered as a permutation of the products in the sample of products. Considering each of the sampled learning orders, model learning is performed using the $\text{PL}^*$ method and the non-adaptive learning method. The following methods are used to generalize the results by catering for the following random exogenous variables in the learning process:
\begin{enumerate}
    \item Using random orders for the input alphabet
    \item Using random orders for the initial prefixes
    \item Using random orders for the initial suffixes
\end{enumerate}
The total amount of the learning cost metrics for each learning method is calculated for each combination of random values. The total value of each metric for each learning order is calculated using the sum of the values of that metric for learning the model of all products in that order. The amount of metrics in the $\text{PL}^*$ method and the non-adaptive learning method are compared using the one-sided paired sample T-test.

\subsubsection{The Effect of Learning Order}\label{learning_order}
To evaluate and quantify the effect of learning order on the efficiency of the $\text{PL}^*$ method, for each subject SPL two learning orders are considered: one learning order with a high learning efficiency and one with a low learning efficiency. To obtain these learning orders, the results of the previous experiments are sorted in ascending order based on the total number of resets, the total number of input symbols and the total number of rounds, respectively. In the resulting table, the first row corresponds to the order with the highest learning efficiency and the last row corresponds to the order with the lowest learning efficiency among the orders tested. Considering these learning orders, the product models are learned using the $\text{PL}^*$ method and the total amount of metrics are calculated. This experiment is repeated 50 times for each order. The amount of metrics for these learning orders are compared using the non-paired T-test.

\section{Results}\label{sec:results}

In this section, we first present the results of the experiments performed to evaluate the efficiency of the proposed adaptive learning method in comparison to the non-adaptive algorithm. Then, we show how different orderings of the products affect the total cost of learning in the adaptive algorithm. Finally, we discuss the obtained results and the threats to their validity.
To make the diagrams clearer, the scale of each diagram is adjusted according to the values in that diagram.
In this section, the average and standard deviation values of the number of resets and the number of input symbols are rounded to the nearest whole number.
For the metrics shown in Tables 1 to 7, the standard deviation in the non-adaptive learning method is zero and hence, is not shown in the table.
To highlight the amount of improvement made by the $\text{PL}^*$ method, ``improvement percentage'' is defined.
For each learning cost metric, if $m$ is its value in the $\text{PL}^*$ method and $m'$ is its value in the non-adaptive learning method, the improvement percentage is calculated as $(1-\frac{m}{m'})*100\%$.
For each specific learning cost metric, if the improvement percentage is positive, it means that using the $\text{PL}^*$ method improves learning efficiency in terms of that metric.
On the other hand, if the improvement percentage of some metric is negative, it shows that using the $\text{PL}^*$ method reduces learning efficiency in terms of that metric.

\subsection{Comparing the learning methods (RQ1-RQ3)}\label{e1}

\begin{figure}[!ht]
    \centering
    \includegraphics[trim=0 0.65cm 0 0,width=0.98\linewidth]{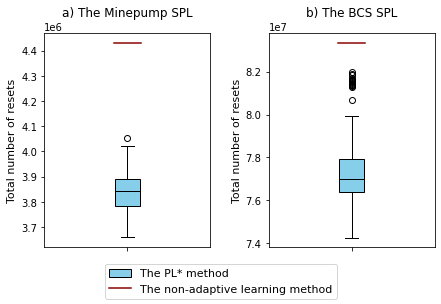}
    \caption{Distribution of the total number of resets}
    \label{fig:figure_4}
\end{figure}

\begin{figure}[!ht]
    \centering
    \includegraphics[trim=0 0.65cm 0 0,width=0.98\linewidth]{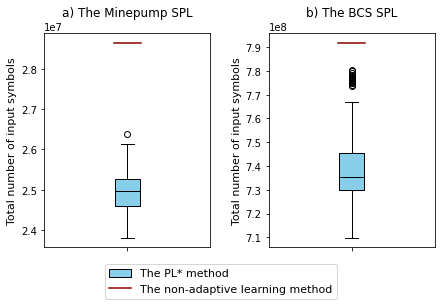}
    \caption{Distribution of the total number of input symbols}
    \label{fig:figure_5}
\end{figure}

In this experiment, the total amount of the learning cost metrics is calculated for 200 random orders using the $\text{PL}^*$ method and the non-adaptive learning method. In the non-adaptive learning method, the values of the learning cost metrics are exactly the same for all orders tested, obviously.  Figure \ref{fig:figure_4} shows the distribution of the total number of resets for the subject SPLs. The distribution of the total number of input symbols is shown in Figure \ref{fig:figure_5}. In these figures, the blue diagrams show the box-plots of the metrics for the $\text{PL}^*$ method. The values of metrics for the non-adaptive learning method are represented by the horizontal line on top of each box plot.

The values of the efficiency metrics in the $\text{PL}^*$ method and the non-adaptive learning method are compared using the one-sided paired sample T-test. Tables \ref{tab:total_rounds}, \ref{tab:total_resets}, and \ref{tab:total_input_symbols} summarize the results of these tests for the number of rounds, the total number of resets, and the total number of input symbols, respectively. In the table for each metric, the ``Ratio'' column shows the ratio of the value of that metric in the $\text{PL}^*$ method to the value of the same metric in the non-adaptive learning method; Ratio values are rounded to three decimal places.

Table \ref{tab:total_rounds} shows that in the Minepump SPL, the use of the $\text{PL}^*$ method reduces the number of learning rounds by about 39\% compared to the non-adaptive learning method. In the BCS SPL, the number of rounds decreases by about 23\%. The results of one-sided paired sample T-tests show that in the tested SPLs, the number of learning rounds in the $\text{PL}^*$ method is significantly less than that of the non-adaptive learning method ($p\text{-value}<0.01$).

\begin{table*}
\caption{The total number of rounds in the $\text{PL}^*$ method and the non-adaptive learning method}
\label{tab:total_rounds}\vspace{-0.35cm}
\centering
{\small 
\begin{tabular}{|c|c|c|c|c|c|}
    \hline
    \multirow{2}{*}{SUL} &
    \multicolumn{2}{|c|}{$\text{PL}^*$ method} &
    \multicolumn{1}{|c|}{Non-adaptive learning method} &
    \multicolumn{1}{|c|}{Improvement} &
    \multicolumn{1}{|c|}{$p$-value} \\
    \cline{2-4}
    & Average & Standard deviation & Average & percentage & (one-sided paired T-test) \\
    \hline
    The Minepump SPL & 18.005 & 1.167 & 30.000 & +39.9\% & $2.845\text{e-}204$\\
    \hline
    The BCS SPL & 16.910 & 0.998 & 22.000 & +23.1\% & $7.034\text{e-}145$\\
    \hline
\end{tabular}
}
\end{table*}

Another metric which is evaluated in these experiments, is the total number of resets, which is the sum of the number of reset operations of the MQs and the EQs. 
Table \ref{tab:total_resets} shows the results of these experiments for the total number of resets.
In the Minepump SPL, the total number of resets in the $\text{PL}^*$ method is approximately 13\% lower than that of the non-adaptive learning method. In the BCS SPL, the amount of the reduction in the number of resets is approximately 7\%. The calculated $p$-values show that the total number of resets in the $\text{PL}^*$ method is significantly less than the amount of this metric in the non-adaptive learning method. Therefore, in the tested SPLs, using the $\text{PL}^*$ method reduces the total number of queries in the learning process.

\begin{table*}
\caption{The total number of resets in the $\text{PL}^*$ method and the non-adaptive learning method}
\label{tab:total_resets}\vspace{-0.35cm}
\centering
{\small
\begin{tabular}[h]{|c|c|c|c|c|c|}
    \hline
    \multirow{2}{*}{SUL} &
    \multicolumn{2}{|c|}{$\text{PL}^*$ method} &
    \multicolumn{1}{|c|}{Non-adaptive learning method} &
    \multicolumn{1}{|c|}{Improvement} &
    \multicolumn{1}{|c|}{$p$-value} \\
    \cline{2-4}
    & Average & Standard deviation & Average & percentage & (one-sided paired T-test) \\
    \hline
    The Minepump SPL & 3,838,078 & 74,075 & 4,429,400 & +13.3\% & $1.095\text{e-}182$\\
    \hline
    The BCS SPL & 77,339,830 & 1,594,173 & 83,332,932 & +7.1\% & $7.259\text{e-}120$\\
    \hline
\end{tabular}
}
\end{table*}

Another effective factor in the efficiency of the model learning algorithms is the length of queries. To estimate this parameter, the total number of input symbols can be used, which is the sum of the input symbols used in MQs and in the implementation of EQs \cite{DBLP:journals/cacm/Vaandrager17}. 
Table \ref{tab:total_input_symbols} shows that using the $\text{PL}^*$ method reduces the total number of input symbols by about 12\% in the Minepump SPL and by about 6\% in the BCS SPL compared to the non-adaptive learning method. 
The above results show that the total number of input symbols in the $\text{PL}^*$ method is significantly less than that of the non-adaptive learning method.

\begin{table*}
\caption{The total number of input symbols in the $\text{PL}^*$ method and the non-adaptive learning method}
\label{tab:total_input_symbols}\vspace{-0.35cm}
\centering
{\small
\begin{tabular}[h]{|c|c|c|c|c|c|}
    \hline
    \multirow{2}{*}{SUL} &
    \multicolumn{2}{|c|}{$\text{PL}^*$ method} &
    \multicolumn{1}{|c|}{Non-adaptive learning method} &
    \multicolumn{1}{|c|}{Improvement} &
    \multicolumn{1}{|c|}{$p$-value} \\
    \cline{2-4}
    & Average & Standard deviation & Average & percentage & (one-sided paired T-test) \\
    \hline
    The Minepump SPL & 24,950,092 & 465,514 & 28,637,112 & +12.8\% & $5.103\text{e-}182$\\
    \hline
    The BCS SPL & 739,258,253 & 14,835,751 & 791,674,093 & +6.6\% & $7.150\text{e-}115$\\
    \hline
\end{tabular}
}
\end{table*}

Therefore, in these experiments, the use of $\text{PL}^*$ method increases the learning efficiency in terms of the number of learning rounds, the total number of resets and the total number of input symbols. 
The number of resets and input symbols are evaluated for MQs and EQs separately.

Tables \ref{tab:MQ_resets} and \ref{tab:MQ_input_symbols} summarize the results of the experiments for the MQ resets and the MQ input symbols, respectively. 
In the Minepump SPL, using the $\text{PL}^*$ method increases the number of MQ resets by about 7\%. In the BCS SPL, the increase in the number of MQ resets is approximately 18\%.
The $\text{PL}^*$ method increases the number of MQ input symbols by about 9\% in the Minepump SPL and by approximately 22\% in the BCS SPL.

\begin{table*}
\caption{The number of MQ resets in the $\text{PL}^*$ method and the non-adaptive learning method}
\label{tab:MQ_resets}\vspace{-0.35cm}
\centering
{\small
\begin{tabular}[h]{|c|c|c|c|c|}
    \hline
    \multirow{2}{*}{SUL} &
    \multicolumn{2}{|c|}{$\text{PL}^*$ method} &
    \multicolumn{1}{|c|}{Non-adaptive learning method} &
    \multicolumn{1}{|c|}{Improvement} \\
    \cline{2-4}
    & Average & Standard deviation & Average & percentage \\
    \hline
    The Minepump SPL & 78,846 & 1,193 & 73,937 & -6.7\% \\
    \hline
    The BCS SPL & 757,186 & 43,247 & 642,412 & -17.9\% \\
    \hline
\end{tabular}
}
\end{table*}

\begin{table*}
\caption{The number of MQ input symbols in the $\text{PL}^*$ method and the non-adaptive learning method}
\label{tab:MQ_input_symbols}\vspace{-0.35cm}
\centering
{\small
\begin{tabular}[h]{|c|c|c|c|c|}
    \hline
    \multirow{2}{*}{SUL} &
    \multicolumn{2}{|c|}{$\text{PL}^*$ method} &
    \multicolumn{1}{|c|}{Non-adaptive learning method} &
    \multicolumn{1}{|c|}{Improvement} \\
    \cline{2-4}
    & Average & Standard deviation & Average & percentage \\
    \hline
    The Minepump SPL & 433,967 & 7,513 & 401,613 & -8.1\% \\
    \hline
    The BCS SPL & 5,826,720 & 351,470 & 4,804,082 & -21.3\% \\
    \hline
\end{tabular}
}
\end{table*}

Tables \ref{tab:EQ_resets} and \ref{tab:EQ_input_symbols} show the experiment results for the EQ resets and the EQ input symbols, respectively. 
In the Minepump SPL, using the $\text{PL}^*$ method reduces the number of EQ resets by about 13\%. In the BCS SPL, the amount of reduction in the number of EQ resets is approximately 7\%. 
The $\text{PL}^*$ method decreases the number of EQ input symbols by about 13\% in the Minepump SPL and by about 6\% in the BCS SPL. 

 \begin{table*}
\caption{The number of EQ resets in the $\text{PL}^*$ method and the non-adaptive learning method}
\label{tab:EQ_resets}\vspace{-0.35cm}
\centering
{\small
\begin{tabular}[h]{|c|c|c|c|c|}
    \hline
    \multirow{2}{*}{SUL} &
    \multicolumn{2}{|c|}{$\text{PL}^*$ method} &
    \multicolumn{1}{|c|}{Non-adaptive learning method} &
    \multicolumn{1}{|c|}{Improvement} \\
    \cline{2-4}
    & Average & Standard deviation & Average & percentage \\
    \hline
    The Minepump SPL & 3,759,232 & 73,731 & 4,355,463 & +13.6\% \\
    \hline
    The BCS SPL & 76,582,644 & 1,575,454 & 82,690,520 & +7.3\% \\
    \hline
\end{tabular}
}
\end{table*}

\begin{table*}
\caption{The number of EQ input symbols in the $\text{PL}^*$ method and the non-adaptive learning method}
\label{tab:EQ_input_symbols}\vspace{-0.35cm}
\centering
{\small
\begin{tabular}[h]{|c|c|c|c|c|}
    \hline
    \multirow{2}{*}{SUL} &
    \multicolumn{2}{|c|}{$\text{PL}^*$ method} &
    \multicolumn{1}{|c|}{Non-adaptive learning method} &
    \multicolumn{1}{|c|}{Improvement} \\
    \cline{2-4}
    & Average & Standard deviation & Average & percentage \\
    \hline
    The Minepump SPL & 24,516,124 & 463,540 & 28,235,499 & +13.1\% \\
    \hline
    The BCS SPL & 733,431,533 & 14,740,136 & 786,870,011 & +6.7\% \\
    \hline
\end{tabular}
}
\end{table*}

The results of the experiments show that the $\text{PL}^*$ method can improve the learning efficiency in terms of the total number of rounds, resets and input symbols. The $\text{PL}^*$ method increases the number of MQs. This adaptive learning method reduces the total cost of learning by reducing the number of EQs. Tables \ref{tab:MQ_resets} and \ref{tab:EQ_resets} show that in the subject SPLs, the number of EQ resets is at least one order of magnitude higher than the number of MQ resets. Similarly, Tables \ref{tab:MQ_input_symbols} and \ref{tab:EQ_input_symbols} show that the number of EQ input symbols is at least one order of magnitude higher than the number of MQ input symbols. These results indicate that the impact of EQs on the total cost of learning is much greater than the effect of MQs. Therefore, in both subject SPLs, the $\text{PL}^*$ method increases the total learning efficiency.

In the $\text{PL}^*$ method, the observation table of the product under learning is initialized using sequences from the previously learned models which are defined in its alphabet. Therefore, it makes the initial observation table more similar to the final observation table (the observation table after learning). As a result, the $\text{PL}^*$ method can decrease the number of rounds. This learning method is suitable for SPLs which are complex enough that the model learning of some of their products requires more than one round. We have not yet evaluated the effect of caching on the learning methods. 

\subsection{The Effect of Learning Order (RQ4)}

To evaluate the effect of learning order on the $\text{PL}^*$ algorithm, from each subject SPL, two learning orders are selected: one order with a high learning efficiency (order 1) and one order with a relatively low learning efficiency (order 2), as explained in \ref{learning_order}. Considering these orders, the model learning is performed using the $\text{PL}^*$ method. The experiment is repeated 50 times for each of these learning orders. The results of order 1 and order 2 are compared using the two-sided unpaired T-test.

In the Minepump SPL, the total number of learning rounds in all repetitions of this experiment is 15 for order 1 and 22 for order 2. However, to learn these models using the non-adaptive learning method, 30 rounds are required. As mentioned earlier, the efficiency of the non-adaptive learning method does not depend on the learning order of products. In the BCS SPL, the total number of rounds in the $\text{PL}^*$ method is 16 for order 1 and 18 for order 2, while the number of rounds in the non-adaptive learning method is 22.

\begin{figure}[!ht]
    \centering
    \includegraphics[trim=0 0.65cm 0 0,width=0.98\linewidth]{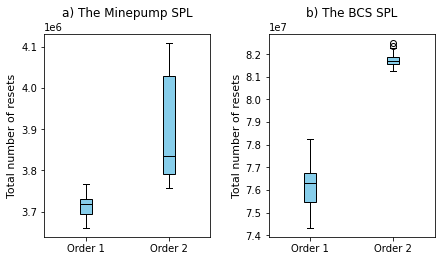}
    \caption{Distribution of the total number of resets in the experimented learning orders}
    \label{fig:o_resets}
\end{figure}

Table \ref{tab:table_4} shows the results of these experiments on the total number of resets. In the Minepump SPL, the average of the total number of resets is 3714556.240 for order 1, while the value of this metric is 3898983.160 for order 2. The $p$-value of the two-sided unpaired T-test is $2.875\text{e-}14$. In the BCS SPL, the average of the total number of resets is 76182731.480 for order 1 and 81718889.180 for order 2. In this experiment, the calculated $p$-value is $3.281\text{e-}42$. Therefore, in these experiments, the total number of resets in order 1 is significantly different from that of order 2. Figure \ref{fig:o_resets} shows the distribution of the total number of resets in the experimented learning orders.

\begin{table*}
\caption{The effect of product learning order on the total number of resets in the $\text{PL}^*$ method}
\label{tab:table_4}\vspace{-0.35cm}
\centering
{\small
\begin{tabular}[h]{|c|c|c|c|c|c|}
    \hline
    \multirow{2}{*}{SUL} &
    \multicolumn{2}{|c|}{Learning order 1} &
    \multicolumn{2}{|c|}{Learning order 2} &
    \multicolumn{1}{|c|}{$p$-value} \\
    \cline{2-5}
    & Average & Standard deviation & Average & Standard deviation & (two-sided unpaired T-test) \\
    \hline
    The Minepump SPL & 3,714,556 & 25,498 & 3,898,983 & 124,018 & $2.875\text{e-}14$\\
    \hline
    The BCS SPL & 76,182,731 & 978,964 & 81,718,889 & 269,506 & $3.281\text{e-}42$\\
    \hline
\end{tabular}
}
\end{table*}

\begin{figure}[!ht]
    \centering
    \includegraphics[trim=0 0.65cm 0 0,width=0.98\linewidth]{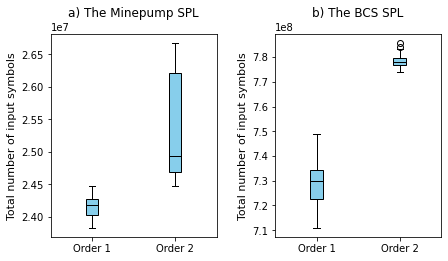}
    \caption{Distribution of the total number of input symbols in the experimented learning orders}
    \label{fig:o_input_symbols}
\end{figure}

Table \ref{tab:table_5} summarizes the results of the experiments on the total number of input symbols. In the Minepump SPL, the average of the total number of input symbols is 24166589.700 for order 1 and 25374188.260 for order 2. The $p$-value of the two-sided unpaired T-test is $1.421\text{e-}14$. In the BCS SPL, the average of the total number of input symbols is 728779454.060 for order 1 and 778246372.540 for order 2. The calculated $p$-value in this experiment is $2.071\text{e-}40$. Therefore, the total number of input symbols in order 1 is significantly different from the same metric in order 2. Figure \ref{fig:o_input_symbols} shows the distribution of the total number of input symbols in the experimented orders.

\begin{table*}
\caption{The effect of product learning order on the total number of input symbols in the $\text{PL}^*$ method}
\label{tab:table_5}\vspace{-0.35cm}
\centering
{\small
\begin{tabular}[h]{|c|c|c|c|c|c|}
    \hline
    \multirow{2}{*}{SUL} &
    \multicolumn{2}{|c|}{Learning order 1} &
    \multicolumn{2}{|c|}{Learning order 2} &
    \multicolumn{1}{|c|}{$p$-value} \\
    \cline{2-5}
    & Average & Standard deviation & Average & Standard deviation & (two-sided unpaired T-test) \\
    \hline
    The Minepump SPL & 24,166,590 & 160,544 & 25,374,188 & 796,102 & $1.421\text{e-}14$\\
    \hline
    The BCS SPL & 728,779,454 & 9,355,222 & 778,246,373 & 2,470,848 & $2.071\text{e-}40$\\
    \hline
\end{tabular}
}
\end{table*}

The results show that in the $\text{PL}^*$ method, the order of learning the products can affect the efficiency of model learning in the SPL context.

\subsubsection{How to Determine the Order of Learning}

As mentioned earlier, the order of learning products can affect the efficiency of the $\text{PL}^*$ method.
In this experiment, the use of parameter $D$ to determine the product learning order in the $\text{PL}^*$ method is evaluated.
Using the equation \ref{eq:1}, the parameter $D$ is calculated for all 200 learning orders in Experiment \ref{e1}. The Pearson correlation coefficient $r$ between the parameter $D$ and the learning cost metrics and its $p$-value is calculated.

\begin{figure}[!ht]
    \centering
    \includegraphics[trim=0 0.6cm 0 0,width=0.98\linewidth]{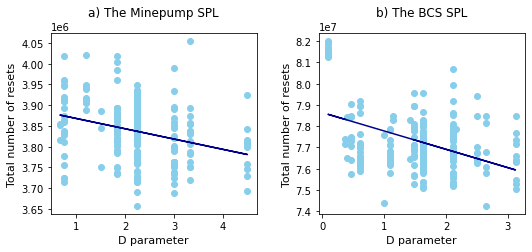}
    \caption{Diagram of the total number of resets vs. the parameter $D$}
    \label{fig:learning_orders_resets}
\end{figure}

Table \ref{tab:r_1} summarizes the results of these experiments for the total number of resets. The Pearson correlation coefficient between the parameter $D$ and the total number of resets is -0.305 for the Minepump SPL ($p$-value $=1.127\text{e-}05$) and -0.430 for the BCS SPL ($p$-value $=2.183\text{e-}10$). 
Figure \ref{fig:learning_orders_resets} shows the diagrams of the total number of resets vs. parameter $D$ and its regression line.

\begin{table}
\caption{The Pearson correlation coefficient between the parameter $D$ and the total number of resets}
\label{tab:r_1}\vspace{-0.35cm}
\centering
{\small
\begin{tabular}[h]{|c|c|c|}
    \hline
    SUL & $r$ & $p$-value \\
    \hline
    The Minepump SPL & -0.305 & $1.127\text{e-}05$ \\
    \hline
    The BCS SPL & -0.430 & $2.183\text{e-}10$ \\
    \hline
\end{tabular}
}
\end{table}

\begin{figure}[!ht]
    \centering
    \includegraphics[trim=0 0.6cm 0 0,width=0.98\linewidth]{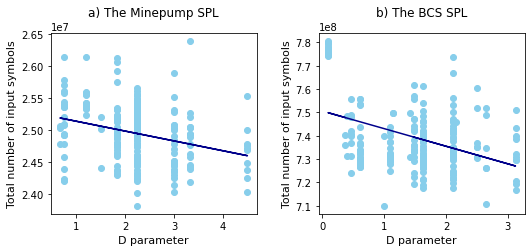}
    \caption{Diagram of the total number of input symbols vs. the parameter $D$}
    \label{fig:learning_orders_input_symbols}
\end{figure}

The correlation coefficient between the parameter $D$ and the total number of input symbols and its $p$-value is summarized in Table \ref{tab:r_2}. The Pearson correlation coefficient between the parameter $D$ and the total number of input symbols is -0.301 for the Minepump SPL ($p$-value $=1.484\text{e-}05$) and -0.404 for the BCS SPL ($p$-value $=2.988\text{e-}09$). 
The diagrams of the total number of input symbols against the parameter $D$ and its regression line are shown in Figure \ref{fig:learning_orders_input_symbols}.

\begin{table}
\caption{The Pearson correlation coefficient between the parameter $D$ and the total number of input symbols}
\label{tab:r_2}\vspace{-0.35cm}
\centering
{\small
\begin{tabular}[h]{|c|c|c|}
    \hline
    SUL & $r$ & $p$-value \\
    \hline
    The Minepump SPL & -0.301 & $1.484\text{e-}05$ \\
    \hline
    The BCS SPL & -0.404 & $2.988\text{e-}09$ \\
    \hline
\end{tabular}
}
\end{table}

The above experiments show that the order of learning the products can affect the efficiency of the $\text{PL}^*$ method. 
In these experiments, it is observed that the total learning efficiency usually increases if the number of new non-mandatory features that are added simultaneously is small. Using this observation, the parameter $D$ is defined as a heuristic to find an order which increases the efficiency of learning.
Experimental results show that there is a mild negative correlation between the value of $D$ and the total number of resets. There is also a mild negative correlation between the value of $D$ and the total number of input symbols. Therefore, it is possible to determine the proper order for learning a subset of products using the $\text{PL}^*$ method.
However, for all learning orders tested, the $\text{PL}^*$ method is more efficient than the non-adaptive learning method.

\subsection{Threats to Validity} 
Because the $\text{PL}^*$ method has been tested on a small number of case studies, the results may be biased according to the characteristics of the subject systems.
This is a threat to generalization of our result. To mitigate this threat, we plan to test the $\text{PL}^*$ method on more subject systems.
Evaluating the $\text{PL}^*$ method in case studies featuring the evolution of behavior both in space and time is a way to test more and larger subject systems (considering evolution in time will enrich our set of case studies and make our method applicable to a larger problem space). 
We also see a threat to the generalization of the product ordering heuristic due to our limited set of subject systems; with a large set of case studies, we see a possibility of (statistically) learning the optimal order as well.

The use of a particular T-wise sampling algorithm (with $\text{T}=3$, based on \cite{DBLP:journals/ese/DamascenoMS21}) may pose another threat for the generalization of our results. We plan to extend our results by considering other values for T and more advanced sampling algorithms \cite{DBLP:conf/splc/VarshosazATRMS18}.   

We captured the random exogenous variables involved in our experiments, such as the order of alphabet symbols and prefixes in learning, and minimized their threats to the validity of the results by taking a large sample. This threat is sufficiently mitigated for the current experiment and we did not observe any significant influence of these random variables in our results to plan a further mitigation. 

To minimize the threats to conclusion validity, we opted for the most general statistical tests in our experiment design: for comparing the learning methods, we used one-sided paired sample T-tests. For comparing and evaluating the learning orders, we started off with an unpaired sample T-test but strengthened the results by using the Pearson correlation coefficient for measuring the correlation of learning efficiency with the value of the parameter $D$.

\section{Conclusion}\label{sec:conc}
In this paper, we presented an adaptive model-learning approach that reuses the learned information about the behavior of products while covering the variability space. It has been shown through an empirical evaluation on two subject systems that our proposed adaptive approach significantly outperforms the standard model learning approach based on Angluin's L$^*$ algorithm. For our comparison, we have used the number of resets and the total number of input symbols, as well as the number of equivalence and membership queries. Additionally, we studied the role of product ordering in learning efficiency and provide a heuristic through defining a parameter that was shown to correlate with learning efficiency for our subject systems. 

Performing more experimental evaluation with other subject systems is among our priorities for future work. We plan to extend our technique to a family-based learning process by extending the learning data structures to ones annotated with feature expressions. Other model learning techniques have been proposed recently, which can be extended to the adaptive and family-based setting following the same recipe. 
It was observed that the randomness in the order of prefixes and suffixes affects the efficiency of the $\text{PL}^*$ method, while it does not affect the efficiency of the non-adaptive learning method. Evaluating the effect of randomness of the order of prefixes and suffixes on the efficiency of the $\text{PL}^*$ method is another line of our future research works.

\paragraph{Acknowledgements.} The work of Mohammad Reza Mousavi was partially supported by the UKRI Trustworthy Autonomous Systems Node in Verifiability, Grant Award Reference EP/V026801/2.

\bibliographystyle{ACM-Reference-Format}
\bibliography{splc2022}
\end{document}